\documentclass[a4paper,11pt]{article}
\pdfoutput=1 

\usepackage{jheppub} 

\usepackage[T1]{fontenc} 

\newif\iftopo

\topofalse
\topotrue

\def\bra#1{\mathinner{\langle{#1}|}}
\def\ket#1{\mathinner{|{#1}\rangle}}
\def\braket#1{\mathinner{\langle{#1}\rangle}}

\newcommand{\su}{\uparrow}
\newcommand{\sd}{\downarrow}

\DeclareMathOperator{\tr}{tr}

\title{\boldmath Entanglement Entropy in Pure $Z_2$ Gauge Lattices}

\author[1]{M. Hategan}

\affiliation[1]{University of California Davis, Davis CA 95616, U.S.A.}

\emailAdd{hategan@ucdavis.edu}

\abstract{We show that the Hilbert space of physical states on a pure $Z_2$ gauge lattice in $1 + 1$ and $2 + 1$ dimensions is geometrically separable if the fundamental physical degrees of freedom are taken to be the plaquettes. This results in a physical entanglement entropy that is not affected by gauge fixing. We introduce a lattice model that is physically equivalent to the original and whose entanglement entropy, calculated using link degrees of freedom, is the same as the entanglement entropy calculated using physical states with the addition of a constant boundary term. We also show that, for non-physical gauge link states, entanglement entropy quantifies constraints between gauge choices in plaquettes adjacent to the boundary.}

\begin{document} 
\maketitle
\flushbottom

\section{Introduction}
\label{sec:intro}

The geometric separation of the Hilbert space of physical states in lattice gauge theories appears to have some difficulties. It seems that the consensus is that a factorization based on assigning individual links to different regions is not possible without sacrifices (see, e.g.,~\cite{buividovich_entanglement_2008,Casini:2013rba,Radicevic:2014kqa,Aoki:2015bsa,Ghosh:2015iwa,Donnelly:2011hn}). Several solutions have been proposed. In~\cite{buividovich_entanglement_2008}, the Hilbert space is extended with additional states at the boundary. This violates gauge invariance at the boundary. Casini et al.~\cite{Casini:2013rba} note that gauge fixing leads to a separable space, but that the value of the entanglement entropy is affected by the particular choice of gauge fixing condition. In~\cite{Radicevic:2014kqa}, ``buffer zones'', which are additional plaquettes, are added to the model at the boundary. The splitting of links at the boundary is proposed in~\cite{Donnelly:2011hn} and further analyzed in~\cite{Casini:2013rba,Ghosh:2015iwa}. Existing literature exposes two classes of problems: 1) different choices in the separation of the link space lead to different entanglement entropies and 2) modifications of the original lattice model which address the first type of problem introduce the issue of faithfulness to the physics of the initial model.

\section{Gauge invariant states}
\label{sec:states}

\begin{figure}[tbp]
\centering
\includegraphics[width=.70\textwidth]{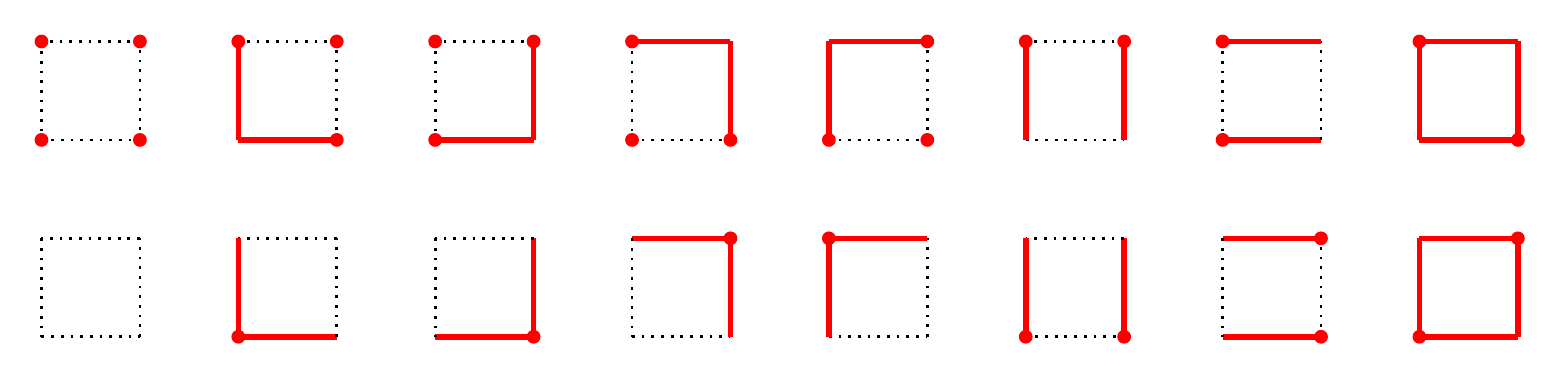}
\caption{\label{fig:plaquette-gt} Gauge transformations on a $Z_2$ plaquette. The gauge transformations happen at the marked vertices. The solid links are links that are affected by the gauge transformation. Transformations in the same column are equivalent.}
\end{figure}

Consider a lattice with $Z_2$ plaquettes, a set of link operators $U_i$ acting on link states $\ket{u_i}$, and a basis which diagonalizes $U_i$: $U_i\ket{\su_i} = +\ket{\su_i}, U_i\ket{\sd_i} = -\ket{\sd_i}$. Simple gauge transformations at a vertex are operators that flip all the links directly connected to that vertex. Given any plaquette in a lattice, a simple gauge transformation will flip either two or none of its links. Complex gauge transformations are obtained by applying different combinations of simple gauge transformations. For any plaquette in the lattice, a complex gauge transformation will always flip an even number of links. By explicit enumeration or other means, one can see that the size of the gauge orbit for an isolated plaquette is $8$. In other words, given a link configuration, $7$ more distinct configurations can be derived from it using only gauge transformations (Figure~\ref{fig:plaquette-gt}). Since the total number of possible configurations of $4$ $Z_2$ links is $16$, we can conclude that there are two states per plaquette that do not mix under gauge transformations. We call these states ``physical''.

While the number of possible physical degrees of freedom observable for an individual plaquette is set, this does not mean that they are always independent. In the one spatial dimensional case, as well as in two 
spatial dimensions with free boundary conditions, the physical degrees of freedom are independent. In the latter case, when periodic boundary conditions are considered, or in $3(+1)$ dimensions, this is no longer true, and certain combinations of physical states on plaquettes are not possible. This results in constraints or entanglement between physical states on plaquettes that are due to the geometry of the model. These constraints, however, do not change the basic assumption of the number of physical states observable on an individual plaquette. We will focus on the cases in which the physical degrees of freedom are independent since it is easier to analyze but still considered problematic in literature. Additionally, we adopt a temporal gauge, which sets all the links in the time direction to one.

In general, state functionals on a lattice are analytic functions of link variables, $l_i$, which take values in $\{-1, 1\}$ corresponding to the states $\{\ket{\su}, \ket{\sd}\}$. Since $l_i^2 = 1$, we are left with only the choice of whether to include a particular link or not in polynomial terms. A basis for functionals is:

\begin{equation}
	\Psi[l_i] = \prod_{p_i \in \{0, 1\}} l_i^{p_i}.
\end{equation}

The inner product on this space is:

\begin{equation}
	\braket{\Psi_1|\Psi_2} = \dfrac{1}{2^N}\sum_{l_i \in \{-1, 1\}} \Psi^*_1[l_i]\Psi_2[l_i],
\end{equation}

with $i \in \{1 \ldots N\}$ and $N$ being the total number of links. The space of functionals is isomorphic to the space of quantum states in which link values are replaced by link operators acting on link states:

\begin{equation}
	\braket{\Psi_1|\Psi_2} = \dfrac{1}{2^N}\sum_{\substack{i = 1, N \\ u_i \in \{\sd, \su\}}} \bra{u_1 \ldots u_N} \Psi^*_1[U_i]\Psi_2[U_i] \ket{u_1 \ldots u_N}.
\end{equation}

For example, given the functional $\Psi[l] = \dfrac{1}{2}\left(a(1 + l) + b(1 - l)\right)$, where $l$ is $Z_2$ valued, the coefficients for the two possible values of $l$ are:

\begin{align}
	\Psi[-1] &= \dfrac{1}{2}\left(0 + 2b\right) = b\\
	\Psi[1] &= \dfrac{1}{2}\left(2a + 0\right) = a.
\end{align}

When $l$ is replaced by the operator $U$ acting on spin states $\ket{u}$, we have:

\begin{align}
	\Psi[U]\ket{\sd} &= \dfrac{1}{2}\left(a(1 + U) + b(1 - U)\right)\ket{\sd} = a\ket{\sd} \\
	\Psi[U]\ket{\su} &= \dfrac{1}{2}\left(a(1 + U) + b(1 - U)\right)\ket{\su} = b\ket{\su}.
\end{align}

The state represented by the above functional is $\ket{\Psi} = a\ket{\sd} + b\ket{\su}$. We will use the two formulations interchangeably. Gauge transformations act on links:

\begin{equation}
	l_i \rightarrow g_{ia} l_i g_{ib}^{-1},
\end{equation}

where $g_{ia}$ and $g_{ib}$ are gauge group elements associated with the endpoints of the link. An arbitrary gauge transformation acts on an arbitrary link polynomial as follows:

\begin{equation}
	l_{i_1} l_{i_2} ... l_{i_n} \rightarrow g_{i_1a} l_{i_1} g_{i_1b}^{-1} g_{i_2a} l_{i_2} g_{i_1b}^{-1} \ldots g_{i_na} l_{i_n} g_{i_nb}^{-1}.
\end{equation}

This polynomial is gauge invariant only if all of the gauge terms cancel out, which can only happen for closed paths (Wilson loops), products of closed paths, or a constant. The smallest loop goes around a single plaquette. This implies that we can use Wilson loop functionals to construct functionals $\Psi[U_i] = \sum c_k W_k[U_i]$ that result in gauge invariant states:

\begin{equation}
	\ket{\Psi} = \sum_{u_i \in \{\sd, \su\}} \Psi[U_i] \ket{u_1 \ldots u_N} = \sum_{u_i \in \{\sd, \su\}} \sum_{k} c_k W_k[U_i] \ket{u_1 \ldots u_N},
\end{equation}

where $W_k[U_i]$ are any subset of the Wilson loop polynomials, including the identity. These states are gauge invariant because they assign the same coefficients to all microstates $\ket{u_1 \ldots u_N}$ related by gauge transformations.

The two physical states of a plaquette can be expressed in a basis constructed using Wilson loop functionals:

\begin{align}
	P &= U_1 U_2 U_3 U_4 \\
	P^2 &= 1 \\
	P_+ &= \dfrac{1}{\sqrt{2}} \left(1 + P\right) \label{eq:pp}\\
	P_- &= \dfrac{1}{\sqrt{2}} \left(1 - P\right) \label{eq:pm}\\
	\ket{\Psi_+} &= P_+\ket{0} \\
	\ket{\Psi_-} &= P_-\ket{0},
\end{align}

where

\begin{align}
	\ket{0} &= \dfrac{1}{4}\sum_{u_i \in \{\sd, \su\}} \ket{u_1 u_2 u_3 u_4} \\ \nonumber
	        &= \dfrac{1}{4}\left[\ket{\sd\sd\sd\sd} + \ket{\sd\sd\sd\su} + \ldots + \ket{\su\su\su\su}\right].
\end{align}

The normalization constant is chosen such that $\braket{0|0} = 1$. We note that:

\begin{align}
	\bra{0}P\ket{0} &= 0 \\
	P_+^2 &= \dfrac{1}{2} (1 + P)(1 + P) = \dfrac{1}{2}(1 + 2P + 1) = \sqrt{2}P_+ \\
	P_-^2 &= \sqrt{2}P_- \\
	P_-P_+ &= \dfrac{1}{2} (1 - P)(1 + P) = \dfrac{1}{2}(1 - 1) = 0.
\end{align}

The link variables are real, therefore $P_+^\dagger = P_+$ and $P_-^\dagger = P_-$. The two states are orthonormal:

\begin{align}
	\bra{0}P_+ P_-\ket{0} &= 0 \\
	\bra{0}P_+ P_+\ket{0} &= \sqrt{2}\bra{0}P_+\ket{0} = \bra{0}1 + P\ket{0} = \braket{0|0} + \bra{0}P\ket{0} = 1.
\end{align}

This can also be seen by looking at the microstates:

\begin{align}
	P_+\ket{0} &= \dfrac{1}{2\sqrt{2}}\left[\ket{\su\su\su\su} + \ket{\sd\sd\su\su} + \ldots + \ket{\sd\sd\sd\sd}\right] \text{(even \# of }\su\text{)} \\
	P_-\ket{0} &= \dfrac{1}{2\sqrt{2}}\left[\ket{\sd\su\su\su} + \ket{\su\sd\su\su} + \ldots + \ket{\sd\sd\sd\su}\right] \text{(odd \# of }\su\text{)}.
\end{align}

In this basis, the plaquette operator, $P$, is diagonal:

\begin{align}
	P P_- \ket{0} &= \dfrac{1}{\sqrt{2}} P \left(1 - P\right) \ket{0} = \dfrac{1}{\sqrt{2}} (P - 1)\ket{0} = -P_-\ket{0} \\
	P P_+ \ket{0} &= \dfrac{1}{\sqrt{2}} P \left(1 + P\right) \ket{0} = \dfrac{1}{\sqrt{2}} (P + 1)\ket{0} = +P_+\ket{0}.
\end{align}

In other words, it is always possible to determine whether the number of links in a plaquette microstate is odd or even, and the two situations correspond to the two distinct physical states of a plaquette. A state on a full lattice can be expressed as a linear combination of basis states which are products of operators $P_{\pm}$ for each plaquette, starting from the full lattice ``ground state'':

\begin{align}
	\ket{0} &= \dfrac{1}{\sqrt{2^N}}\sum_{u_i \in \{\sd, \su\}} \ket{u_1 \ldots u_N} \\
	\ket{\Phi_p} &= \prod_{x} P_{p_x,x}\ket{0} \\
	\ket{\Psi} &= \sum_p c_p \ket{\Phi_p},
	\label{eq:full-state}
\end{align}

where $p = (p_1, \ldots, p_M), p_x \in \left\{-,+\right\}$ are $2^M$ dimensional vectors that index the basis vectors of the physical states, $M$ is the number of plaquettes, and $c_p$ are coefficients satisfying $\sum c_p^2 = 1$. The basis states are orthonormal:

\begin{equation}
	\braket{\Phi_p|\Phi_{p'}} = \bra{0}\prod_x P_{p_x, x} P_{p'_x, x}\ket{0} = \prod_x \delta_{p_x,p'_x}.
	\label{eq:basis-states-on}
\end{equation}

This is apparent for two reasons. First, any product of the form $P_{+,x} P_{-,x} = \dfrac{1}{2} (1 - P_x^2)$ is zero, since $P_x^2 = 1$. Therefore, $p_x$ must equal $p'_x$ for all $x$ in order to get a non-zero result. Second, if all $p_x = p'_x$, then the left hand side of~\ref{eq:basis-states-on} reduces to:

\begin{equation}
	\bra{0}\prod_x (1 \pm P_{x})\ket{0} = \braket{0|0} + \bra{0}\sum f(P_x)\ket{0},
\end{equation}

where $f(P_x)$ are various terms that contain at least one link operator. Such terms vanish, since they are anti-symmetric with respect to $\ket{0}$. In order to obtain a reduced density matrix, we can divide the set of plaquettes into two regions, $A$ and $\bar{A}$ such that $p= (q_1, \ldots, q_{M_A}, \bar{q}_1, \ldots, \bar{q}_{M - M_A}) = q \oplus \bar{q}$. The reduced density matrix, $\rho_A$, for some state, $\ket{\Psi}$, is then obtained by summing over the degrees of freedom in $\bar{A}$:

\begin{align}
	\rho_A[\Psi, q, q'] &= \sum_{\bar{q}}\braket{\Phi_{q \oplus \bar{q}}|\Psi}\braket{\Psi|\Phi_{q' \oplus \bar{q}}} \\
		&= \sum_{\bar{q}} 
			\bra{0} 
			\prod_{\substack{x \in A \\ \bar{x} \in \bar{A}}}
			P_{q_x,x} P_{\bar{q}_{\bar{x}}, \bar{x}}
			\ket{\Psi}
			\bra{\Psi}
			\prod_{\substack{x \in A \\ \bar{x} \in \bar{A}}}
			P_{q'_x,x} P_{\bar{q}_{\bar{x}}, \bar{x}}
			\ket{0}.
\end{align}

The entanglement entropy~\cite{Srednicki:1993im} is then:

\begin{equation}
	S_A = \tr\rho_A \ln\rho_A.
\end{equation}

The above representation of states is independent of gauge fixing to the extent that the algebra of the $P_{\pm,x}$ operators is the same and only $\ket{0}$ changes. This leads to a reduced density matrix and an entanglement entropy that are gauge invariant.

One may note that the gauge invariant space is equivalent to a matter spin system with spin degrees of freedom being the plaquettes, where the operator $P$ plays the role of $\sigma_z$. To complete the algebra, we must find $\sigma_x$ and show that its commutator with $\sigma_z$ leads to an operator equivalent to $\sigma_y$. We have at our disposal, the remaining gauge invariant operators, $L_i$, which flip the state of the links $s_i$. First, for an isolated plaquette, we note that:

\begin{align}
	L_i \ket{0} &= L_i \sum_{u_i} \ket{\ldots u_i \ldots} = \sum_{u_i} \ket{\ldots \bar{u_i} \ldots} = \ket{0} \\
	L_i U_i \ket{u_i} &= L_i l_i \ket{u_i} = l_i \ket{\bar{u}_i} = - U_i L_i \ket{u_i} \\
	L_i P \ket{0} &= L_i U_1 U_2 U_3 U_4 \ket{0} = - P L_i \ket{0}.
\end{align}

The effect of $L_i$ on the gauge invariant states of a plaquette is:

\begin{align}
	L_i P_- \ket{0} &= \dfrac{1}{\sqrt{2}}L_i (1 - P) \ket{0} = \dfrac{1}{\sqrt{2}}(L_i + P L_i) \ket{0} = P_+ \ket{0}\label{eq:opLonPstates1} \\
	L_i P_+ \ket{0} &= \dfrac{1}{\sqrt{2}}L_i (1 + P) \ket{0} = \dfrac{1}{\sqrt{2}}(L_i - P L_i) \ket{0} = P_- \ket{0}\label{eq:opLonPstates2}.
\end{align}

\begin{figure}[tbp]
\centering
\includegraphics[width=.70\textwidth]{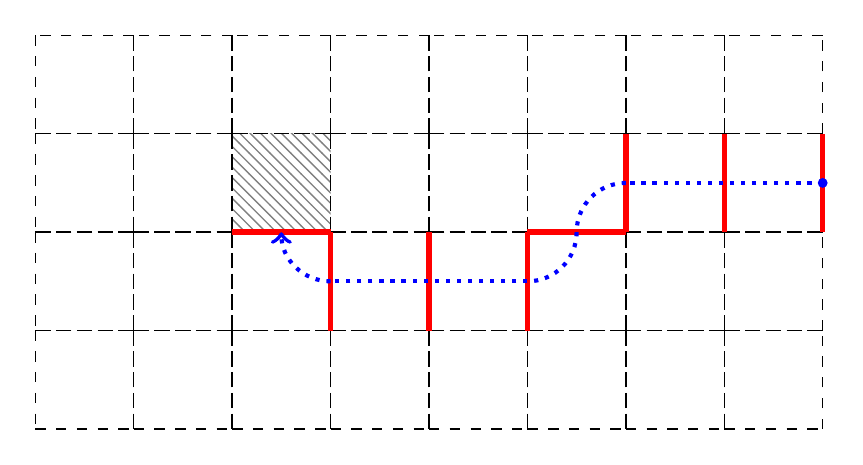}
\caption{\label{fig:worm} Construction of one of the possible plaquette-flip operators. The operator acts on the shaded plaquette and is composed of link flip operators on links that are shown in thick, red, lines. The operator acts as the identity on the physical space of the plaquettes that it ``passes through''.}
\end{figure}

Therefore $L_i$ plays the role of $\sigma_x$ on an isolated plaquette. On a full lattice, links are shared between plaquettes. The $L_i$ operators end up flipping the state of both plaquettes that share a link. However, products of $L_i$ operators that start with a link on an edge plaquette that is not shared by other plaquettes and end at a certain plaquette (Figure~\ref{fig:worm}), act as a $\sigma_x$ for that plaquette, since products of two $L_i$ that act on links of the same plaquette act as the identity on the physical space:

\begin{align}
	L_i L_j P_\pm \ket{0} = L_i P_\mp \ket{0} = P_\pm \ket{0}.
\end{align}

The importance of a lattice with free boundary conditions is apparent here. Nonetheless, in the case of, e.g., periodic boundary conditions, there is a single Mandelstam constraint, and a basis can be chosen such that the state of a single plaquette is linearly dependent on the state of rest of the plaquettes. Such a plaquette can serve as a ``boundary'' with free links where the operators above can terminate. The remaining operator in the spin $\frac{1}{2}$ algebra can be obtained from the commutator of $L_i$ and $P$:

\begin{align}
	\sigma_y = -\dfrac{i}{2} [P, L_i] = -i P L_i.
\end{align}

Its effect on the physical states is as expected:

\begin{align}
	\sigma_y P_-\ket{0} &= -i P L_i P_- \ket{0} = -i P P_+\ket{0} = -i P_+\ket{0} \\
	\sigma_y P_+\ket{0} &= -i P L_i P_+ \ket{0} = -i P P_-\ket{0} = i P_-\ket{0}.
\end{align}

A similar treatment to the entanglement entropy of Abelian lattice gauge theory based on the duality to spin systems can be found in~\cite{Radicevic:2016tlt}.

\iftopo
\section{Topological model and topological entanglement entropy}

According to~\cite{Casini:2013rba}, the wave functional for a topological $Z_2$ lattice gauge model can be taken as:

\begin{align}
	\Psi[l_i] = K \sum_{\Gamma} \prod_{k \in \Gamma} u_k,
\end{align}

where $\Gamma$ is the set of all subsets of links that form strings that are not open. Wilson loop operators form a group under multiplication~\cite{Casini:2013rba}: bigger loops can be written as products of smaller loops. Therefore, the functional above can be written as:

\begin{align}
	\Psi[l_i] = K \sum_{p_x \in \{0, 1\}} \prod_x P_x^{p_x}[l_i],
\end{align}

where $P_x$ are plaquette operators and $x$ indexes all the plaquettes in the lattice. The functional can be factored as follows:

\begin{align}
	\Psi &= K (P_0 + 1) \sum_{p_x \in \{0, 1\}} \prod_{\substack{x \\ x > 0}} P_x^{p_x} \\
		&= K\sqrt{2} P_{+,0} \sum_{p_x \in \{0, 1\}} \prod_{\substack{x \\ x > 0}} P_x^{p_x},
\end{align}

where we used Eq.~\ref{eq:pp} and Eq.~\ref{eq:pm}. The process can be iterated over all $x$ to obtain:

\begin{align}
	\Psi = K' \prod_x P_{+,x}.
\end{align}

This state is equivalent to a spin chain with all spins pointing up, whose entanglement entropy is zero, irrespective of the choice of regions. The topological entanglement entropy~\cite{Kitaev:2005dm} is then trivially zero. This is to be expected, since there are no physical correlations in this state.
\fi

\section{The two-plaquette lattice}

\begin{figure}[tbp]
\centering
\includegraphics[width=.30\textwidth]{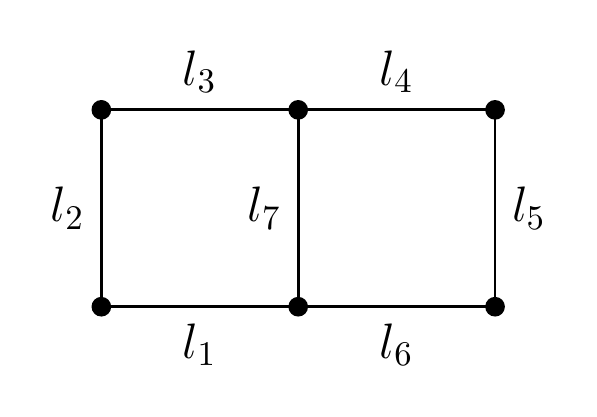}
\caption{\label{fig:simple-lattice} A simple two-plaquette lattice}
\end{figure}

A simple model that can be used to gain some insight into the differences between the entanglement of physical and unphysical states is the two-plaquette $Z_2$ lattice (Figure~\ref{fig:simple-lattice}).

There are two plaquette operators in this model:

\begin{align}
	P_L &= U_1 U_2 U_3 U_7 \\
	P_R &= U_4 U_5 U_6 U_7.
\end{align}

If we define

\begin{align}
	P_{\{L, R\}\pm} &= \dfrac{1}{\sqrt{2}} \left(1 \pm P_{\{L, R\}}\right),
\end{align}

then the basis for physical states that diagonalizes both $P_L$ and $P_R$ is:

\begin{align}
	\ket{\Psi_{--}} &= P_{L-} P_{R-} \ket{0} \\
	\ket{\Psi_{-+}} &= P_{L-} P_{R+} \ket{0} \\
	\ket{\Psi_{+-}} &= P_{L+} P_{R-} \ket{0} \\
	\ket{\Psi_{++}} &= P_{L+} P_{R+} \ket{0}.
\end{align}

\begin{figure}[tbp]
\centering
\includegraphics[width=.24\textwidth]{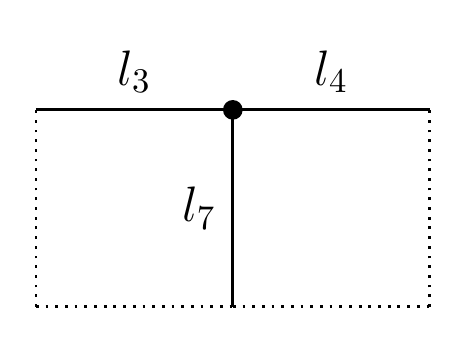}
\caption{\label{fig:simple-lattice-gf} Partial gauge fixing on the simple lattice. The dotted links are set to $1$.}
\end{figure}

We note here that the link-flip operator acting on the shared link, $L_7$, does not make the physical space inseparable. Its effect on the physical space is to flip the physical state of both plaquettes (see Eq.~\ref{eq:opLonPstates1} and Eq.~\ref{eq:opLonPstates2}). It is equivalent to the operator $\sigma_{x, L} \sigma_{x, R}$ acting on two unconstrained spins and it can be written as a tensor product of operators acting on the individual plaquettes/spins.

In order to gain a better understanding of the differences in entanglement entropy between the physical picture and the link basis, we fix the gauge partially as shown in Figure~\ref{fig:simple-lattice-gf}. There is a single gauge transformation possible, which flips all the remaining free links. Consider a state formed by a single basis vector:

\begin{equation}
	\ket{\Psi} = \ket{\Psi_{--}}.
\end{equation}

This corresponds to the state $\ket{\sd\sd}$ in a two spin system and its physical entanglement entropy is zero. The link microstates are:

\begin{equation}
	\ket{\Psi} = P_{L-} P_{R-} \ket{0} = \dfrac{1}{2}(1 - U_3 U_7) (1 - U_4 U_7) \ket{0} = \dfrac{1}{\sqrt{2}}\left[\ket{\su_3\sd_7\su_4} + \ket{\sd_3\su_7\sd_4}\right].
\end{equation}

The microstates are physically indistinguishable, since they differ only by a gauge transformation. However, separating the links space such that $l_4$ is in one region and $l_3, l_7$ are in its complement (the electric center choice in~\cite{Casini:2013rba,Ghosh:2015iwa}), we would have exactly one qubit of entanglement, since a choice of $l_4$ fully determines $l_3$ and $l_7$. Compared to the case of two separate plaquettes, this reflects an entanglement between their respective gauge orbits. This entanglement is purely a matter of geometry and, ignoring dynamics, the distinctions between connected plaquettes and separated plaquettes is the number of distinct gauge transformations that are possible and how the resulting microstates are entangled.

In fact, the link entanglement entropy above remains the same for all physical states of the form

\begin{equation}
	\ket{\Psi'} = \alpha\ket{\Psi_{--}} + \beta\ket{\Psi_{++}},
\end{equation}
with $\alpha^2 + \beta^2 = 1$. The physical entanglement entropy of this state can be anywhere between zero (for either $\alpha = 1$ or $\beta = 1$) and one (for $\alpha = \beta = 1 / \sqrt{2}$). The link microstates are:

\begin{equation}
	\ket{\Psi'} = 
		\dfrac{\alpha}{\sqrt{2}}\left[\ket{\su_3\sd_7\su_4} + \ket{\sd_3\su_7\sd_4}\right] +
		\dfrac{\beta}{\sqrt{2}}\left[\ket{\su_3\su_7\su_4} + \ket{\sd_3\sd_7\sd_4}\right].
\end{equation}

The reduced density matrix obtained by tracing over links $7$ and $4$ is:

\begin{align}
	\rho_{7,4} &= \dfrac{\alpha^2}{2} \ket{\su_3}\bra{\su_3} + 
				  \dfrac{\alpha^2}{2} \ket{\sd_3}\bra{\sd_3} + 
				  \dfrac{\beta^2}{2} \ket{\su_3}\bra{\su_3} + 
				  \dfrac{\beta^2}{2} \ket{\sd_3}\bra{\sd_3} \\
			   &= \dfrac{1}{2}\ket{\su_3}\bra{\su_3} + \dfrac{1}{2} \ket{\sd_3}\bra{\sd_3}.
\end{align}

The resulting entanglement entropy is constant and independent of $\alpha$ and $\beta$. In other words, for the family of states above, the electric center choice leads to a constant entanglement entropy that is independent of the entanglement entropy that two observers could agree on. Numerical simulations\footnote{The simulation code is available at \href{https://github.com/hategan/ee}{https://github.com/hategan/ee}} show that this remains true even without the partial gauge fixing. One can conclude that, in general, an entanglement entropy obtained using non-physical link operators cannot be used to determine the entanglement entropy between physical states.

\section{Modified lattice model}

\begin{figure}[tbp]
\centering
\includegraphics[width=.35\textwidth]{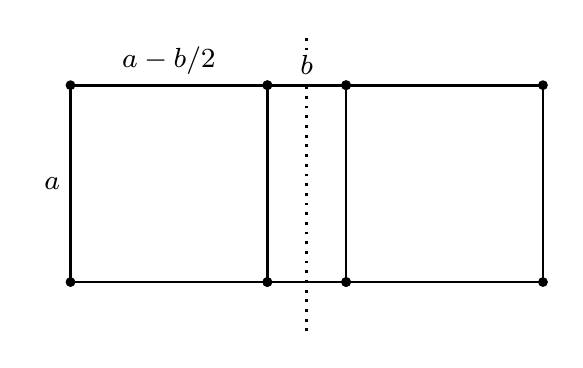}
\caption{\label{fig:modified-model} A modified lattice. The links crossing the boundary have length $b$.}
\end{figure}

The standard lattice model can be modified slightly to allow a more direct separation of links by introducing a ``boundary plaquette'', as shown in Figure~\ref{fig:modified-model}. We assume a continuous Abelian gauge group, such as $U(1)$ and switch to $3 + 1$ dimensions. The model should reproduce the same physics as the original model in the limit of $b \rightarrow 0$. We first consider a non-isotropic plaquette in the $x-y$ plane with a length of $b$ and a height of $a$. We can express the plaquette as a product of links:
\newcommand{\hx}{\hat{x}}
\newcommand{\hy}{\hat{y}}
\begin{equation}
	U_{xy}(n) = U_{x}(n) U_{y}(n + b\hx) U_{x}^\dagger(n + a\hy)U_{y}^\dagger(n).
\end{equation}

Using $U_{x}(n) = \exp(i b A_{x}(n)), U_{y}(n) = \exp(i a A_{y}(n))$~\cite{gattringer}:

\begin{equation}
	U_{xy}(n) = \exp\big(i b A_{x}(n) + i a A_{y}(n + b\hx) - i b A_{x}(n + a\hy) - i a A_{y}(n)\big).
\end{equation}

The gauge fields can also be Taylor-expanded to first order to yield:

\begin{align}
	U_{xy}(n) &= \exp\big(i b A_{x}(n) + i a (1 + b\partial_x) A_{y}(n) - i b (1 + a\partial_y) A_{x}(n) - i a A_{y}(n)\big) \\
				  &\approx \exp\big(i a b (\partial_x A_y(n) - \partial_y A_x(n))\big) \\
				  &= \exp(i a b F_{xy}(n)).
\end{align}

By Taylor-expanding the exponential and taking the real part, we obtain:

\begin{equation}
	\mathrm{Re}\,U_{xy}(n) = 1 - a^2 b^2 F_{xy}^2(n) + O(a^3 b^3),
\end{equation}

therefore:

\begin{equation}
	\mathrm{Re}\,\big(1 - U_{xy}(n)\big) \approx a^2 b^2 F_{xy}^2(n)
\end{equation}

When $a = b$ and summing over the entire space, the factors of $a$ serve as the integration measure in the discretized integral, and, in the continuum limit, we can identify $a^4 F_{\mu\nu}^2$ with $\mathrm{d^4 x} F_{\mu\nu}^2$. This allows us to connect the continuum action with the discretized action:

\begin{equation}
	{d^4 x} \dfrac{1}{4} F_{\mu\nu}^2 = a^4 \dfrac{1}{4} F_{\mu\nu}^2 = \dfrac{1}{4} \mathrm{Re}\,\big(1 - U_{\mu\nu}(n)\big).
\end{equation}

However, for a non-isotropic plaquette, the measure should be $a^3 b$. In other words:

\begin{equation}
	{d^4 x} \dfrac{1}{4} F_{xy}^2 = a^3 b \dfrac{1}{4} F_{xy}^2 = \dfrac{a}{b} a^2 b^2 \dfrac{1}{4}F_{xy}^2 = \dfrac{a}{b} \dfrac{1}{4}\mathrm{Re}\,\big(1 - U_{xy}(n)\big).
\end{equation}

In the limit of $b \rightarrow 0$, and assuming a sufficiently slow varying gauge field such that we do not introduce momentum modes above the lattice cutoff, the links of length $b$ become the identity: $U_x(n) \approx 1$. We can therefore write the boundary plaquette term as:

\begin{equation}
	{d^4 x} \dfrac{1}{4} F_{xy}^2 = \dfrac{a}{b}\dfrac{1}{4} \mathrm{Re}\big(1 - U_y(n + b\hx) U_y^\dagger(n)\big).
\end{equation}

With the large factor of $a / b$, such terms represent dynamical constraints that set all ${U_y(n)}$ equal to ${U_y(n+b\hx)}$. This exercise is not strictly necessary. We could have simply considered two plaquettes separated by a small space and then constrain the adjacent links to be equal. Such a model would reproduce the same physics as the original one.

We can now return to a $Z_2$ model in $2 + 1$ dimensions and compare the entanglement entropy calculated using physical states with the entanglement entropy calculated using links in a separated two-plaquette model with the adjacent link constraint. We first consider an arbitrary physical state and use the spin notation:

\begin{align}
	\ket{\Psi} &= \alpha \ket{\sd\sd} + \beta \ket{\sd\su} + \gamma \ket{\su\sd} + \delta \ket{\su\su} \\
			   &= [\alpha \ket{\sd} + \gamma \ket{\su}] \otimes \ket{\sd} + [\beta \ket{\sd} + \delta \ket{\su}] \otimes \ket{\su}
\end{align}

The reduced density matrix is:

\newcommand{\exabsq}{\ensuremath{\alpha^2 + \beta^2}}
\newcommand{\excdsq}{\ensuremath{\gamma^2 + \delta^2}}
\newcommand{\exacbd}{\ensuremath{\alpha\gamma + \beta\delta}}

\begin{equation}
	\rho_B = \bordermatrix{~         & \bra{\sd} & \bra{\su} \cr
	                       \ket{\sd} & \exabsq    & \exacbd   \cr
	                       \ket{\su} & \exacbd   & \excdsq}.
\end{equation}

The the corresponding physical state in the modified model is:

\newcommand{\bsd}{\Downarrow}
\newcommand{\bsu}{\Uparrow}

\begin{align}
	\ket{\Psi'} &= \dfrac{1}{\sqrt{2}} \big[ 
					\alpha (\ket{\bsd\sd\sd\bsd} + \ket{\bsd\su\su\bsd}) 
					+ \beta (\ket{\bsd\sd\sd\bsu} + \ket{\bsd\su\su\bsu}) \\
					&+ \gamma (\ket{\bsu\sd\sd\bsd} + \ket{\bsu\su\su\bsd}) 
					+ \delta (\ket{\bsu\sd\sd\bsu} + \ket{\bsu\su\su\bsu})
				\big] \\
				&= \dfrac{1}{\sqrt{2}} \big[
				    (\alpha\ket{\bsd\sd} + \gamma\ket{\bsu\sd}) \otimes \ket{\sd\bsd}
				  + (\beta\ket{\bsd\sd} + \delta\ket{\bsu\sd}) \otimes \ket{\sd\bsu} \\
				 &+ (\alpha\ket{\bsd\su} + \gamma\ket{\bsu\su}) \otimes \ket{\su\bsd}
				  + (\beta\ket{\bsd\su} + \delta\ket{\bsu\su}) \otimes \ket{\su\bsu}
				\big],
\end{align}

where $\ket{\bsd x}$ and $\ket{\bsu x}$ stand for all link configurations that result in a physical up or down state, respectively, given a constrained boundary link of $x$. These link configurations represent gauge transformations that do not involve the boundary links. Consequently, these configurations are independent between the two plaquettes and can be factorized. The resulting reduced density matrix is:

\begin{align}
	\rho'_B = \dfrac{1}{2}\bordermatrix{
		~             & \ket{\bsd\sd} & \ket{\bsd\su} & \ket{\bsu\sd} & \ket{\bsu\su} \cr
		\ket{\bsd\sd} & \exabsq       & 0             & \exacbd       & 0             \cr
		\ket{\bsd\su} & 0             & \exabsq       & 0             & \exacbd       \cr
		\ket{\bsu\sd} & \exacbd       & 0             & \excdsq       & 0             \cr
		\ket{\bsu\su} & 0             & \exacbd       & 0             & \excdsq
	}.
\end{align}

This can be written as:

\newcommand{\id}{\mathbb{I}_2}
\begin{align}
	\rho'_B = \dfrac{1}{2} \begin{pmatrix}
		\exabsq & \exacbd \\
		\exacbd & \excdsq
	\end{pmatrix}\otimes\id.
\end{align}

The above expression of $\rho_B'$ is essentially the same as $\rho_B$. Now, assume that $\rho_B$ has the following diagonal form:

\begin{equation}
	\rho_B = \begin{pmatrix}
		\lambda_1 & 0 \\
		0 & \lambda_2
	\end{pmatrix},
\end{equation}

with $\tr\rho_B = \lambda_1 + \lambda_2 = 1$. Then, the physical entanglement entropy is:

\begin{equation}
	S_B = - \lambda_1 \log \lambda_1 - \lambda_2 \log \lambda_2.
\end{equation}

Furthermore, $\rho_B'$ in diagonal form is:

\begin{equation}
	\rho_B' = \dfrac{1}{2}
			\begin{pmatrix}
				\lambda_1 & 0 \\
				0 & \lambda_2 
			\end{pmatrix}
			\otimes\id.
\end{equation}

The link entanglement entropy of the modified model is:

\begin{align}
	S_B' &= -\dfrac{\lambda_1}{2}\log\dfrac{\lambda_1}{2}\times 2 - \dfrac{\lambda_2}{2}\log\dfrac{\lambda_2}{2} \times 2 \\
	     &= -\lambda_1 \log \lambda_1 - \lambda_2 \log \lambda_2 + (\lambda_1 + \lambda_2) \log 2 \\
	     &= S_B + \log 2.
\end{align}

In other words, the link entanglement entropy of the modified model is equal to the physical entanglement entropy with the addition of a constant term due to the constraint between the boundary links.

\section{Separability and algebras}

\begin{figure}[tbp]
\centering
\includegraphics[width=.24\textwidth]{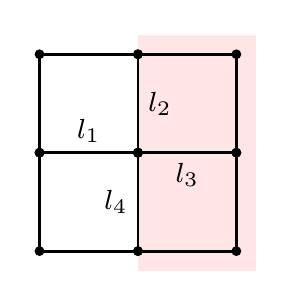}
\caption{\label{fig:four-p-lattice} A four-plaquette lattice. The shaded region on the right is the ``inside'', or $V$. Full gauge freedom is allowed. For simplicity, only some of the links are labeled.}
\end{figure}

In this section we look at the implications of the statements in~\cite{Casini:2013rba} about the non-separability of gauge theories due to the existence of operator constraints that cross region boundaries. The setup is illustrated in Figure~\ref{fig:four-p-lattice}. The operators $L_i$, which act on individual links and flip their value, $L_i\ket{\su_i} = \ket{\sd_i}, L_i\ket{\sd_i} = \ket{\su_i}$, are considered. These operators are gauge invariant. Intuitively, the effect of flipping $l_1$ in Figure~\ref{fig:four-p-lattice} is the same as the effect of flipping all the links emanating from the center vertex, flipping $l_1$ and then flipping all the links emanating from the center vertex back ($L_1 = g L_1 g^{-1}$). However, in the physical space, there exist constraints between the $L_i$ operators. In particular (see eq. 14 in~\cite{Casini:2013rba}):

\begin{equation}
	L_1 L_2 L_3 L_4 = 1.
	\label{eq:gauge-constraint}
\end{equation}

This is not surprising, since flipping all four links connected to the central vertex in Figure~\ref{fig:four-p-lattice} amounts to a gauge transformation, which leaves physical states invariant. The above equation can be equivalently written as:

\begin{equation}
	L_2 L_3 L_4 = L_1^{-1} = L_1.
\end{equation}

If the space is separable, then the sub-spaces are associated with two algebras such that all operators belonging to each sub-algebra can be written as either $O_V \otimes 1_{\bar{V}}$ or $1_V \otimes O_{\bar{V}}$. However, it is argued, since $L_1$, an operator in $\bar{V}$, is equal to $L_2 L_3 L_4$, which is a non-trivial operator in $V$, then such a decomposition of operators is not possible. We note however that the constraint in eq.~\ref{eq:gauge-constraint} is only valid in the physical space. On the other hand, on the Hilbert space of links, flipping $l_1$ leads to an entirely different microstate than flipping $l_2$, $l_3$, and $l_4$. In the physical space of plaquettes in $V$, flipping $l_2$, $l_3$, and $l_4$ is indistinguishable from a gauge transformation. In other words, an observer with access only to observables in $V$, would not be able to see the effects of either $L_1$ or $L_2 L_3 L_4$ and, on the physical state of plaquettes in $V$, both are identity operators.

\section{Conclusion}

In this paper, we show that, for a $1 + 1$ and $2 + 1$ $Z_2$ lattice gauge theory with open boundary conditions, the Hilbert space of physical states is geometrically separable. This separation cannot be achieved by assigning links to regions, in agreement with~\cite{buividovich_entanglement_2008}, since physical constraints induce a physical Hilbert space with different degrees of freedom from those of the link space. Instead, the geometrical separation must be done by assigning plaquettes to regions. The resulting entanglement entropy is, unsurprisingly, independent of gauge fixing conditions. If, instead, a separation based on the non-physical link space is considered, the entanglement entropy reflects an entanglement between gauge degrees of freedom. However, the entanglement entropy obtained from link states is, in general, independent of what could be determined by two observers with access to only physical operators. We show that the two can be reconciled using a modified model that preserves the physics of the original lattice model. In this case, we get $S_{link} = S_{phys} + \log 2$.

\bibliographystyle{JHEP}
\bibliography{ee}
\end{document}